\documentclass[english,aps, prb, twocolumn, superscriptaddress,showpacs,floatfix]{revtex4-1}
\usepackage{lmodern}
\usepackage[T1]{fontenc}
\usepackage[latin9]{inputenc}
\usepackage{textcomp}
\usepackage{graphicx}
\usepackage{bm}
\usepackage{verbatim}
\usepackage{booktabs}
\usepackage{amstext}
\usepackage{amsmath}
\usepackage{color}
\usepackage{upgreek}
\usepackage{multirow}
\usepackage{siunitx}
\usepackage{paralist}

\makeatletter

\newcommand{\lyxmathsym}[1]{\ifmmode\begingroup\def\b@ld{bold}
  \text{\ifx\math@version\b@ld\bfseries\fi#1}\endgroup\else#1\fi}

\@ifundefined{textcolor}{}
{%
 \definecolor{BLACK}{gray}{0}
 \definecolor{WHITE}{gray}{1}
 \definecolor{RED}{rgb}{1,0,0}
 \definecolor{GREEN}{rgb}{0,1,0}
 \definecolor{BLUE}{rgb}{0,0,1}
 \definecolor{CYAN}{cmyk}{1,0,0,0}
 \definecolor{MAGENTA}{cmyk}{0,1,0,0}
 \definecolor{YELLOW}{cmyk}{0,0,1,0}
 \definecolor{darkgreen}{rgb}{0,0.6,0.4}
 }

\usepackage{prettyref}
\newrefformat{tab}{Tab.\,\ref{#1}}
\newrefformat{fig}{Fig.\,\ref{#1}}
\newrefformat{eq}{Eq.\,\textup{(\ref{#1})}}
\newcommand{\LSCoOquart}{La$_{1.75}$Sr$_{0.25}$CoO$_{4}$}

\newcommand{\LSCoOthird}{La$_{1.67}$Sr$_{0.33}$CoO$_{4}$}
\newcommand{\LSCoOfam}{La$_{2-x}$Sr$_{x}$CoO$_{4+\delta}$}

\newcommand{\xicpa}{\xi_{\rm C}^\parallel}
\newcommand{\xicpe}{\xi_{\rm C}^\perp}

\newcommand{\HSP}{\mathcal{H}_{\rm sp}}

\makeatother

\usepackage{babel}

\begin{document}

\title{The Hour-Glass Magnetic Spectrum Arising from a Striped, Cluster Spin Glass Ground State in \LSCoOquart}

\author{S. M. Gaw}
\email[]{s.gaw1@physics.ox.ac.uk}

\affiliation{Department of Physics, University of Oxford, Clarendon Laboratory,
Parks Road, Oxford, OX1 3PU, United Kingdom}

\author{E. C. Andrade}
\author{M. Vojta}
\affiliation{Institut f\"ur Theoretische Physik, Technische Universit\"at Dresden, 01062 Dresden, Germany}

\author{C. D. Frost}
\author{D. T. Adroja}
\affiliation{ISIS Facility, Rutherford Appleton Laboratory, STFC, Chilton, Didcot,
Oxon, OX11 0QX, United Kingdom}

\author{D. Prabhakaran}
\author{A. T. Boothroyd}
\email[]{a.boothroyd@physics.ox.ac.uk}
\affiliation{Department of Physics, University of Oxford, Clarendon Laboratory,
Parks Road, Oxford, OX1 3PU, United Kingdom}

\begin{abstract}
We report inelastic neutron scattering results that reveal an hour-glass magnetic excitation spectrum in \LSCoOquart.  The magnetic spectrum is similar to that observed previously in \LSCoOthird, but the spectral features are broader. We show that the spectrum of \LSCoOquart~can be modelled by the spin dynamics of a system with a disordered cluster spin glass ground state. Bulk magnetization measurements are presented which support the proposed glassy ground state. The observations reiterate the importance of quasi-one-dimensional magnetic correlations and disorder for the hour-glass spectrum, and suggest that disordered spin and charge stripes exist at lower doping in La$_{2-x}$Sr$_{x}$CoO$_{4}$ than previously thought.
\end{abstract}
\maketitle

\section{\label{sec:Introduction} Introduction}

The distinctive \textit{hour-glass} magnetic spectrum first came to prominence in neutron scattering measurements of hole-doped layered copper-oxide superconductors. \cite{AraiPRL1999, BourgesScience2000, TranquadaNature2004, HaydenNature2004, HinkovNature2004, ReznikPRL2004, ChristensenPRL2004, StockPRB2005, LipscombePRL2007, MatsudaPRL2008, XuNatPhys2009} It is characterized by four incommensurate peaks in momentum space, which with increasing energy first disperse inwards towards the square-lattice antiferromagnetic wavevector $(\pi, \pi)$ then disperse outwards again but with a rotation of 45$^{\circ}$ around $(\pi, \pi)$. Various different models have been proposed to account for the hour-glass spectrum in the layered cuprates, from strong coupling models that contain stripe-like correlations to weak coupling models based on itinerant magnetism.\cite{VojtaAdvPhys2009} If spin fluctuations play a role in copper-oxide superconductivity then an understanding of the hour-glass spectrum would be an important step towards a microscopic model.

Qualitatively the same hour-glass spectrum has recently been observed in neutron scattering measurements of certain layered cobalt and manganese oxide insulators.\cite{BoothroydNature2011, UlbrichPRL2012} This suggests that in spite of their very different electronic properties, the superconducting cuprates could harbour the same type of magnetic correlations as found in the layered cobaltates and manganates.  Because the cobaltates and manganates have well localized electrons their magnetic dynamics are relatively easy to understand. Hence, the key requirement for the hour-glass spectrum could be identified as short-range quasi-one-dimensional magnetic correlations, the conditions for which are created in these systems by disordered stripe phases.\cite{BoothroydNature2011, UlbrichPRL2012}

In this paper we examine in more detail the influence of disorder on the magnetic spectrum of striped phases. Stripes are a form of complex order in which charges doped into an antiferromagnet condense into parallel arrays which modulate the magnetic order, such that charge stripes form antiphase domain walls of the background antiferromagnetic order. Even the best-correlated stripe phases have a degree of disorder, as revealed for example by spin-glass features in the magnetization of stripe-ordered cuprates,\cite{ChouPRL1995} nickelates\cite{FreemanPRB2006} and cobaltates.\cite{LancasterArXiv2013} The amount of disorder varies with doping, and it is of interest to investigate the concomitant changes in the hour-glass spectrum.

 The La$_{2-x}$Sr$_x$CoO$_4$ family exhibits nearest-neighbour (nn) antiferromagnetic (AFM) order at $x=0$ (Refs.~\onlinecite{YamadaPRB1989, BabkevichPRB2010}) and robust checkerboard charge ordering of Co$^{2+}$ and Co$^{3+}$ at $x=0.5$ (Refs.~\onlinecite{ZaliznyakPRL2000, HelmePRB2009}). A phase with short-range stripe order has been reported for $0.3\leq x<0.5$ (Ref.~\onlinecite{CwikPRL2009}), and the hour-glass magnetic spectrum was observed in a sample with $x=0.33$ (Ref.~\onlinecite{BoothroydNature2011}). The main features of the hour-glass spectrum could be reproduced by a cluster spin glass model developed for period-3 stripes.\cite{AndradePRL2012}

Here we present inelastic neutron scattering (INS) measurements of La$_{1.75}$Sr$_{0.25}$CoO$_4$ ($x=0.25$). This compound is near the border between the AFM and stripe ordered phases proposed in Ref.~\onlinecite{CwikPRL2009}, where the ground state is likely to be strongly influenced by competing phases.  We find that the magnetic spectrum has an hour-glass shape consistent with period-4 stripes, and that the stripes have a higher degree of disorder than the period-3 stripes present in La$_{1.67}$Sr$_{0.33}$CoO$_4$.  We extend the cluster glass model\cite{AndradePRL2012} to period-4 stripes, and show that the model qualitatively describes the main features of the observed spectrum of La$_{1.75}$Sr$_{0.25}$CoO$_4$.

\section{\label{sec:Experimental Details} Experimental Details}

A single crystal of \LSCoOquart\ with mass \SI{14.3}{\gram} was grown by the optical floating-zone method. Initially, polycrystalline \LSCoOquart\ was prepared from La$_2$O$_3$, SrCO$_3$ and Co$_3$O$_4$ (\textgreater 99.99\% purity) by solid-state reaction. The starting materials were reacted in air at \SI{1200}{\degreeCelsius} for \SI{48}{\hour} then reground and sintered in air at \SI{1225}{\degreeCelsius} for \SI{48}{\hour}. No impurity phases could be detected in the product by x-ray powder diffraction. The powder was then pressed into rods and sintered in air at \SI{1250}{\degreeCelsius} for \SI{24}{\hour}. The crystal growth was performed in a four-mirror image furnace in flowing high purity argon at a growth speed of 2 mm h$^{-1}$ with counter-rotation of the feed and seed rods at 25 rpm. Previous growth experiments on \LSCoOfam\ compounds have shown that it is difficult to achieve oxygen stoichiometry, with most as-grown crystals having an excess of oxygen when $x \lesssim 0.3$. To achieve stoichiometry the as-grown crystal underwent an anneal in a reducing atmosphere of CO$_2$:CO at \SI{850}{\degreeCelsius} for \SI{12}{\hour}.

Sample characterization data were obtained by several different techniques. Electron Probe Microanalysis (EPMA) performed with a Jeol JXA-8600 provided estimates of the cation ratios. X-ray powder diffraction performed on a X'Pert PRO PANalytical diffractometer was used to check phase purity and for basic structure analysis. Finally, magnetic susceptibility and thermo-remnant magnetization measurements were performed with a superconducting quantum interference device (SQUID) magnetometer (Quantum Design).

Neutron scattering spectra were recorded on the MAPS and MERLIN time-of-flight chopper spectrometers at the ISIS facility. The MAPS spectrometer was used with relatively high incident neutron energies ($E_i$) of \SIlist{80;120;300}{\meV} to survey the spectrum over a wide range of energy, $E$. The energy resolution on MAPS was approximately 5\% of $E_i$ at $E=0$, decreasing slightly with increasing $E$. The MERLIN spectrometer was subsequently used with E$_i$=\SI{20}{\meV} to measure the lower energy region of the spectrum with higher resolution, and to measure the spectrum up to 50\,meV with an optimized incident energy of \SI{60}{\meV}. In both experiments, the samples were mounted in a $^4$He closed-cycle refrigerator (CCR) and aligned with the $c$-axis parallel to the incident neutron beam. In this fixed orientation, the intensity was mapped as a function of $E$ and wave vector ${\bf Q}=(H\times 2\pi/a,K\times 2\pi/a,L\times 2\pi/c)$, where $a$ and $c$ are tetragonal lattice parameters, and the out-of-plane wave vector component $L$ varies with $E$. Previous measurements on La$_{2-x}$Sr$_x$CoO$_4$ have shown that for energies greater than a few meV the magnetic correlations between the CoO$_2$ layers are negligible,\cite{BoothroydNature2011} and so the dispersion in the out-of-plane direction can be neglected. This means that the dispersion in the $(H,K)$ plane can be measured directly when the incident neutron beam is perpendicular to the layers.

The magnetic scattering intensity is described by the partial differential cross-section, which in the dipole approximation is given by,\cite{Squires_NeutronScattering}
\begin{equation}
\frac{\partial ^{2}\sigma}{\partial\Omega \partial E}=\frac{k_{\rm f}}{k_{\rm i}}\Big(\frac{\gamma r_0}{2}\Big)^{2}f^2(Q) \sum_{\alpha}(1-\hat{Q}_{\alpha}^2) S^{\alpha\alpha}({\bf Q},E),
\label{eq:S(Q,E)}
\end{equation}

here $k_{\rm i}$ and $k_{\rm f}$ are the incident and final neutron wave vectors, respectively, $(\gamma r_0/2)^2=72.8$ mb, $f(Q)$ is the magnetic form factor for Co$^{2+}$, and $(1-\hat{Q}_{\alpha}^2)$ is the orientation factor. $\hat{Q}_{\alpha}=Q_{\alpha}/Q$ is the $\alpha$ component of the unit vector parallel to $\bf Q$. The scattering function $S^{\alpha\alpha}({\bf Q},E)$ describes the magnetic correlations between the $\alpha$ components of the magnetic moments ($\alpha= x, y, z$ in Cartesian coordinates) and is dependent on ${\bf M}({\bf Q})$, the Fourier transform of the magnetization,
\begin{equation}
S^{\alpha\alpha}({\bf Q},E)=\sum_{j}|\langle j|M^{\alpha}({\bf Q})|0\rangle |^{2} \delta(E - E_j({\bf Q})),
\label{eq:Saa}
\end{equation}
where $\left|0\right\rangle$ is the ground state and $\left|j\right\rangle $ is an excited state with energy $E_{j}$.

The raw data were corrected for detector efficiency, for the time-independent background, and for the $k_{\rm f}/k_{\rm i}$ factor in the differential cross-section, Eq.~(\ref{eq:S(Q,E)}), and measurements of a standard vanadium sample at each incident energy were used to normalize the spectra and place them on an absolute intensity scale. All measurements were made at  $T=6$\,K, the base temperature of the CCR.

\section{\label{sec:Results} Results}

\subsection{\label{Structural and chemical analysis}Structural and Chemical Analysis}

\begin{figure}
\includegraphics[clip,width=\columnwidth]{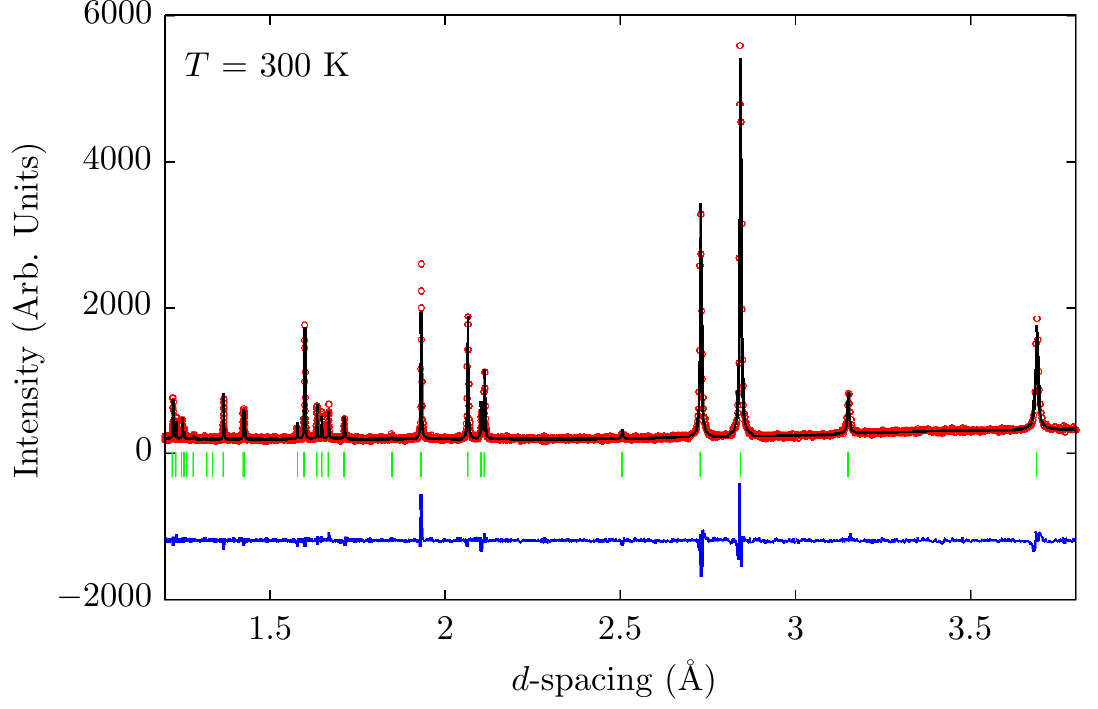}
\caption{(Color online) Powder diffraction pattern of \LSCoOquart\ with Rietveld profile refinement of the tetragonal structure at room temperature. Red circles show the measured data points, the black line is the profile refinement, the green bars show the positions of the refined peaks, and the blue line is the residual (the difference between the data points and refinement). The pattern shows no evidence of impurity phases.}
\label{fig:powder_diffraction}
\end{figure}

\begin{table}
\centering
	\begin{tabular}{ c | c | c c c }
	Nominal doping & 2$\zeta$ & & $x$ & $\delta$ \\
	\hline
	\multirow{3}{*}{ $x=0.25$} & \multirow{3}{*}{0.2240(3)} & Weighed & 0.25 & -0.0130(1) \\
	 & \multirow{3}{*}{} & EPMA & 0.238(4) & -0.007(2) \\
	& \multirow{3}{*}{} & X-Ray & 0.22(3) & 0.00(2) \\
	\end{tabular}
\caption{\label{tab:calc_delta_values} Compositional analysis of \LSCoOfam. The Sr content ($x$) is determined in three different ways: (i) from weighed starting materials, (ii) from EPMA, and (iii) from X-ray occupancy refinement. The parameter $\zeta$ is determined from the magnetic ordering vector ${\bf Q}_{\rm m} = (0.5, 0.5) \pm (\zeta, \zeta)$ observed by neutron diffraction. The oxygen excess ($\delta$) is calculated from the relation $2\zeta = x + 2\delta = n_h$ which is applicable to an ideal stripe structure with a total number of holes $n_{\rm h}$ and one hole per site in the charge stripes.  All values are per formula unit of \LSCoOfam.}
\end{table}

Figure~\ref{fig:powder_diffraction} shows the X-ray powder diffraction pattern recorded at room temperature from the single crystal sample of \LSCoOquart. Additionally, a profile fit from a Rietveld refinement of the data performed with the {\scshape FullProf} suite of programs is presented.\cite{FullProf} The crystal structure of \LSCoOquart\ is described by the space group $I4/mmm$, and the tetragonal unit-cell parameters were refined as $a=3.86348(7)\AA$ and $c=12.6193(3)\AA$ at \SI{300}{\kelvin}. No evidence for additional phases in the sample could be found in the diffraction data.

Knowledge of the Sr and O content of the sample is important to confirm the doping in the system. By assuming the total occupancy of the La/Sr site to be stoichiometric we determined the Sr content ($x$) from EPMA and (separately) from refinement of the X-ray diffraction data. These values are presented in Table~\ref{tab:calc_delta_values}. The more precise value is from EPMA, although the quoted error does not include the possibility of a small systematic error in the calibration.  The results in Table~\ref{tab:calc_delta_values} indicate a very slight deficiency in Sr, but are close to the nominal composition of $x=0.25$. The oxygen is expected to be near stoichiometry due to the annealing process performed post-growth. An indirect check of the oxygen content is presented in Section~\ref{subsect:Magnetic Diffraction} which confirms this expectation. Hereafter, we will assume the nominal composition \LSCoOquart.

\subsection{Magnetic Characterization}

\begin{figure}
\includegraphics[width=\columnwidth]{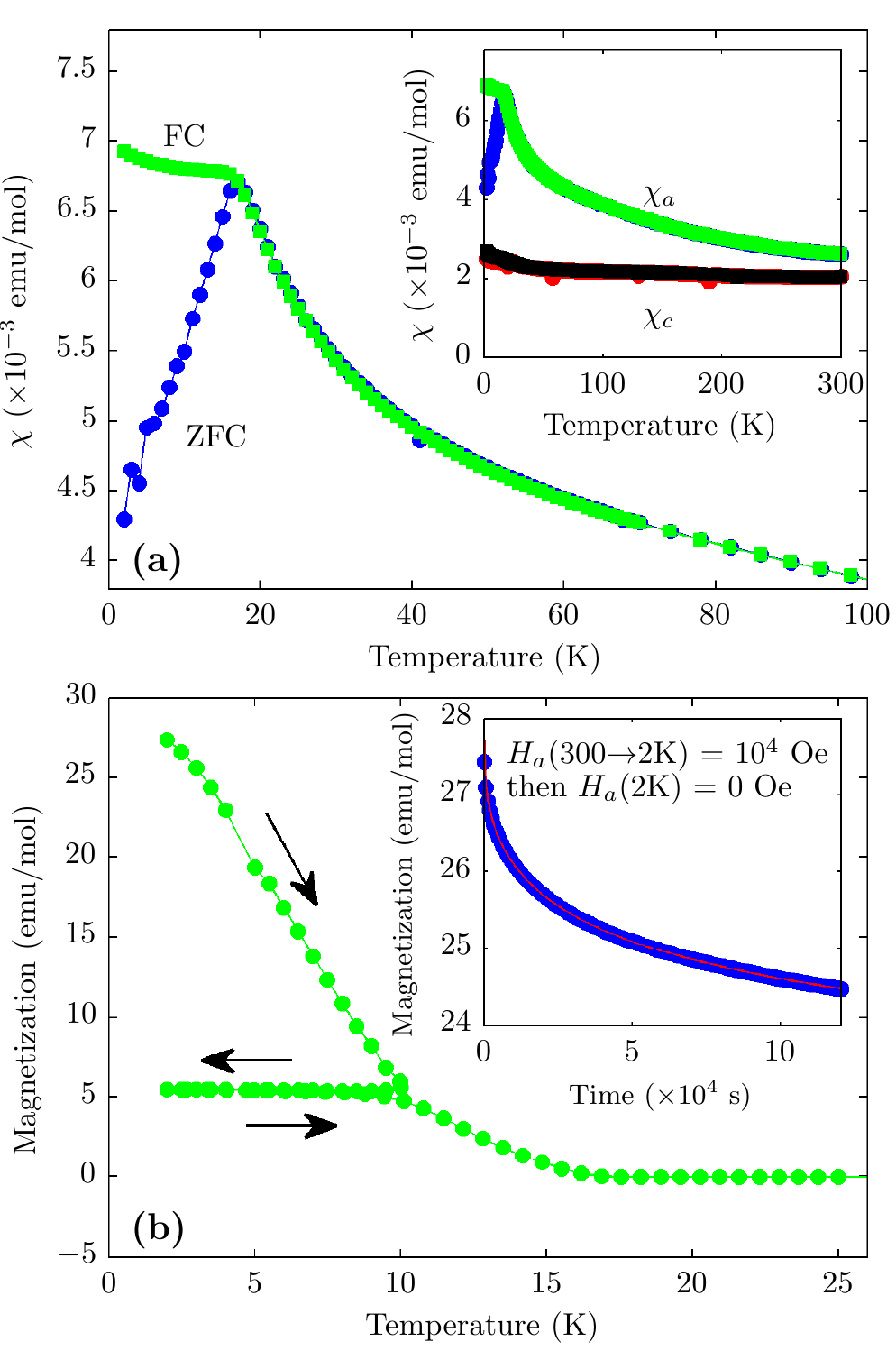}
\caption{(Color online) Susceptibility and magnetization of \LSCoOquart. (a) Temperature dependence of the magnetic susceptibility measured with a field of \SI{1000}\,Oe applied parallel to the $a$-axis. The insert shows the anisotropy in the magnetic susceptibility for fields parallel to the $a$- and $c$-axes. (b) Memory effect observed in the thermo-remnant magnetization. The insert shows the measured decay of TRM with time (blue points) and a fitted stretched exponential (red line), as described in the text.}
\label{fig:SQUID_data}
\end{figure}

Figure~\ref{fig:SQUID_data}(a) shows the susceptibility of \LSCoOquart\ with the measuring field $\bf{H}$ applied parallel to the $a$-axis. Below \SI{18}{\kelvin} there is a splitting of the zero-field-cooled (ZFC) and field-cooled (FC) susceptibilities indicative of a spin freezing transition. The feature in the ZFC measurement at \SI{5}{\kelvin} is an artefact of the magnetometer. The insert to Fig.~\ref{fig:SQUID_data}(a) shows the FC and ZFC susceptibilities for temperatures between 2\,K and 300\,K and measuring fields along the $a$- and $c$-axes. The susceptibility is strongly anisotropic at low temperatures with $\chi_a > \chi_c$, and is qualitatively similar to that of \LSCoOthird\ (Ref.~\onlinecite{BoothroydNature2011}) although $\chi_c$ has a weaker temperature variation in \LSCoOquart.

Several irreversibility effects are seen in the thermo-remnant magnetization (TRM) of \LSCoOquart, shown in Fig.~\ref{fig:SQUID_data}(b). To obtain this data the sample was first cooled to \SI{2}{\kelvin} in a field of \SI{e4} Oe, then the field was removed and the temperature varied in the sequence 2$\rightarrow$10$\rightarrow$2$\rightarrow$\SI{30} K. Overall, the TRM decays with increasing temperature. However, as the temperature sweep is reversed from 10$\rightarrow$\SI{2} K the TRM remains constant. As the sample is warmed again, the TRM remains constant until rejoining the initial decay trend at \SI{10} K. Measurements of the relaxation of the TRM are presented in Fig.~\ref{fig:SQUID_data}(b), insert. The sample was cooled to \SI{2}{\kelvin} in an applied field of \SI{e4} Oe, after which the field was removed and the TRM measured at regular time intervals over about 36 hours. A characteristic decay in the TRM can be seen in the data.

The magnetometry results for \LSCoOquart\ exhibit two of the hallmarks of spin glass behavior. The first is a spin freezing transition in the susceptibility, as seen in Fig.~\ref{fig:SQUID_data}(a). This can be understood as a glassy freezing in of the induced magnetic order (in the FC case) or lack of induced order (in the ZFC case) at \SI{18}{\kelvin}.\cite{Fischer_SpinGlasses}. Furthermore, the pronounced memory effect shown in Fig.~\ref{fig:SQUID_data}(b) illustrates the melting of a frozen induced magnetic order. The second spin-glass hallmark is the characteristic decay trend seen in Fig.~\ref{fig:SQUID_data}(b) insert. This decay was fit to a stretched exponential of the form $M(t)=M_{0}\exp\{-\alpha t^{(1-n)}\}+M_{\rm{bgd}}$, and the value of the stretching exponent $n$ was found to be $n=0.6171(8)$. This exponent approaches that of an ideal spin glass, $n=2/3$ (Ref.~\onlinecite{CampbellPRB1988}). Similar results have been found for the lightly-doped cuprates\cite{ChouPRL1995} and nickelates,\cite{FreemanPRB2006} but with different exponents. Although the precise nature of the spin-glass state seems to be material-dependent, it appears that spin-glass behaviour is a universally shared property associated with the imperfect stripe order of real-world materials.

\subsection{\label{subsect:Magnetic Diffraction}Magnetic Diffraction}

\begin{figure}
\includegraphics[width=\columnwidth]{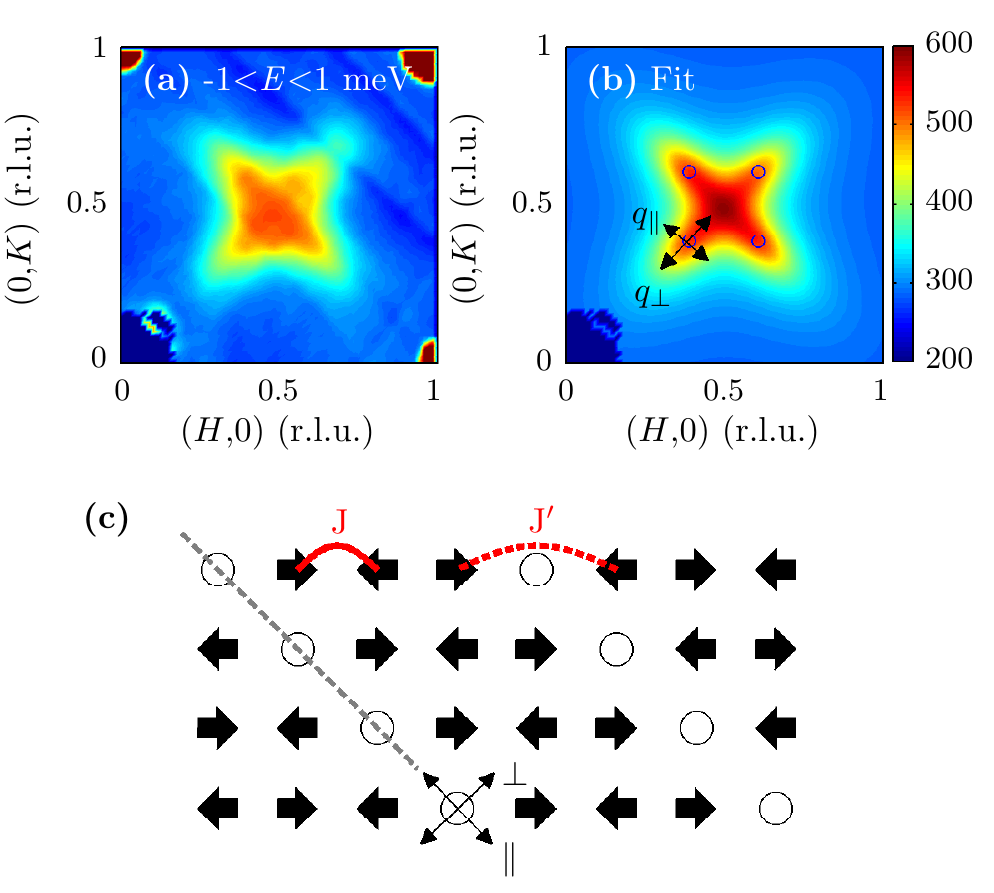}
\caption{\label{fig:elastic_data} (Color online) Short-range magnetic order in La$_{1.75}$Sr$_{0.25}$CoO$_4$. (a) Elastic diffuse neutron scattering intensity in the $(H,K)$ plane measured with $E_{\rm{i}} = 20$\,meV and integrated over the energy range $-1<E<1$\,meV. The intensity is given in units of mb\,sr$^{-1}$\,meV$^{-1}$\,f.u.$^{-1}$. (b) Fit to the elastic data using a pattern of four bi-variant Lorentzian peaks. The peaks are centered on  $\mathbf{Q}_{\rm{m}}=\mathbf{Q}_{\rm{AFM}}\pm(\zeta,\zeta)$ and $\mathbf{Q}_{\rm{AFM}}\pm(\zeta,-\zeta)$. Peak centers are shown as black circles. Peak widths are measured along the $q_{\parallel}$ and $q_{\perp}$ directions (labeled) to determine the correlation lengths parallel and perpendicular to the stripe direction, respectively. (c) Schematic diagram of perfect period-4 stripe order. Open circles denote Co$^{3+}$ ions with $S=0$, while black arrows show Co$^{2+}$ ions with $S=3/2$. The grey dashed line indicates the stripe parallel direction. The nearest-neighbour (solid red) and inter-stripe (dashed red) exchange interactions, $J$ and $J^{\prime}$ respectively, are labeled. This orientation of stripes gives rise to magnetic peaks at $\mathbf{Q}_{\rm{m}}=\mathbf{Q}_{\rm{AFM}}\pm(\zeta,\zeta)$ positions.  Equivalent domains with stripes oriented perpendicular to those depicted in (c) would give rise to  peaks at the $\mathbf{Q}_{\rm{AFM}}\pm(\zeta,-\zeta)$ positions.}
\end{figure}

Figure~\ref{fig:elastic_data}(a) shows the measured elastic neutron scattering in the $(H,K)$ plane in two-dimensional (2D) reciprocal space. The scattering, which is diffuse in character, is centered around the two-dimensional AFM wave vector $\mathbf{Q}_{\rm{AFM}} = (0.5,0.5)$ [ $\equiv (\pi,\pi)$ for a square lattice], and is elongated along the $(1,1)$ and $(-1,1)$ directions.

Previous neutron diffraction studies of hole-doped La$_{2-x}$Sr$_{x}$CoO$_{4}$ compounds with $0.3 < x < 0.5$ have found a four-fold pattern of broad incommensurate magnetic peaks centered on the in-plane wave vectors $\mathbf{Q}_{\rm{m}}=\mathbf{Q}_{\rm{AFM}}\pm(\zeta,\zeta)$ and $\mathbf{Q}_{\rm{AFM}}\pm(\zeta,-\zeta)$.\cite{CwikPRL2009,BoothroydNature2011} The incommensurability $\zeta$ scales with the hole doping $x$, such that $2\zeta = x$. These results have been interpreted as arising from a crystallization of holes into diagonal stripes of Co$^{3+}$ which modulate the background AFM order on the majority Co$^{2+}$ sites --- see Fig.~\ref{fig:elastic_data}(c). Such stripe formation is well established in the isostructural layered nickelates La$_{2-x}$Sr$_{x}$NiO$_{4}$ for $0.15 < x < 0.5$ (Refs.~\onlinecite{ChenPRL1993, TranquadaPRL1994, YoshizawaPRB2000}).

The ``X''-shaped pattern of diffuse scattering seen in Fig.~\ref{fig:elastic_data}(a) is consistent with four broad and overlapping incommensurate peaks surrounding $\mathbf{Q}_{\rm{AFM}}$, and is therefore indicative of short-range stripe order. To quantify this interpretation we fitted the elastic data to an intensity distribution modeled with four bi-variant elliptically-contoured Lorentzian peaks.  This description of the peaks is equivalent to the scattering cross section for anisotropic two-dimensional disorder given by Savici \textit{et al}.\cite{SaviciPRB2007} The centers of the four peaks were shifted by equal amounts $\zeta$ away from $\mathbf{Q}_{\rm{AFM}}$ to fit the incommensurate pattern. The best fit was obtained with $2\zeta = 0.2240(3)$, and the resulting intensity distribution is presented in Fig.~\ref{fig:elastic_data}(b). The calculated distribution is in reasonable agreement with experiment, although the model has a small excess of intensity at $\mathbf{Q}_{\rm{AFM}}$.

The value of the incommensurability parameter $2\zeta = 0.2240$ estimated from this analysis is close to the incommensurability $2\zeta = 0.25$ for ideal period-4 stripes. The difference may be at the same level as the uncertainties in the analysis, but if we naively equate the value of $2\zeta$ with the hole doping then it  suggests a slightly lower doping than is accounted for by the nominal Sr content of $x=0.25$, consistent with the chemical analysis results presented in Table~\ref{tab:calc_delta_values}.  The slightly low value of $2\zeta$ could also be accounted for by a slight deficiency in the oxygen content. The values of the oxygen excess $\delta$ obtained from the relation $2\zeta = x + 2\delta$, which assumes that the stripe period is determined by the total hole concentration from Sr and O, are also given in Table~\ref{tab:calc_delta_values}. The oxygen deficiency values determined this way are very small indeed, which supports our earlier assertion that the crystals are essentially stoichiometric in oxygen.

The elastic diffuse scattering also allows us to quantify the degree of disorder. As the Lorentzian peaks are elliptical, there are two characteristic magnetic correlation lengths, one parallel and the other perpendicular to the stripe direction. The correlation lengths are defined as the inverse of the half-width-at-half-maximum (HWHM) of the incommensurate peaks in the respective directions. These directions are indicated in reciprocal and real space in Figs.~\ref{fig:elastic_data}(b) and (c). From the fit we find $\xi_{\rm{M}}^{\parallel}=7.0\rm{\AA}$ and $\xi_{\rm{M}}^{\perp}=3.5\rm{\AA}$. These values indicate an increase in disorder in La$_{1.75}$Sr$_{0.25}$CoO$_4$ compared with La$_{1.67}$Sr$_{0.33}$CoO$_4$, for which we found $\xi_{\rm{M}}^{\parallel}=10\rm{\AA}$ and $\xi_{\rm{M}}^{\perp}=6.5\rm{\AA}$ (Ref.~\onlinecite{BoothroydNature2011}).

\subsection{Magnetic Excitation Spectrum}

\begin{figure}
\includegraphics[width=\columnwidth]{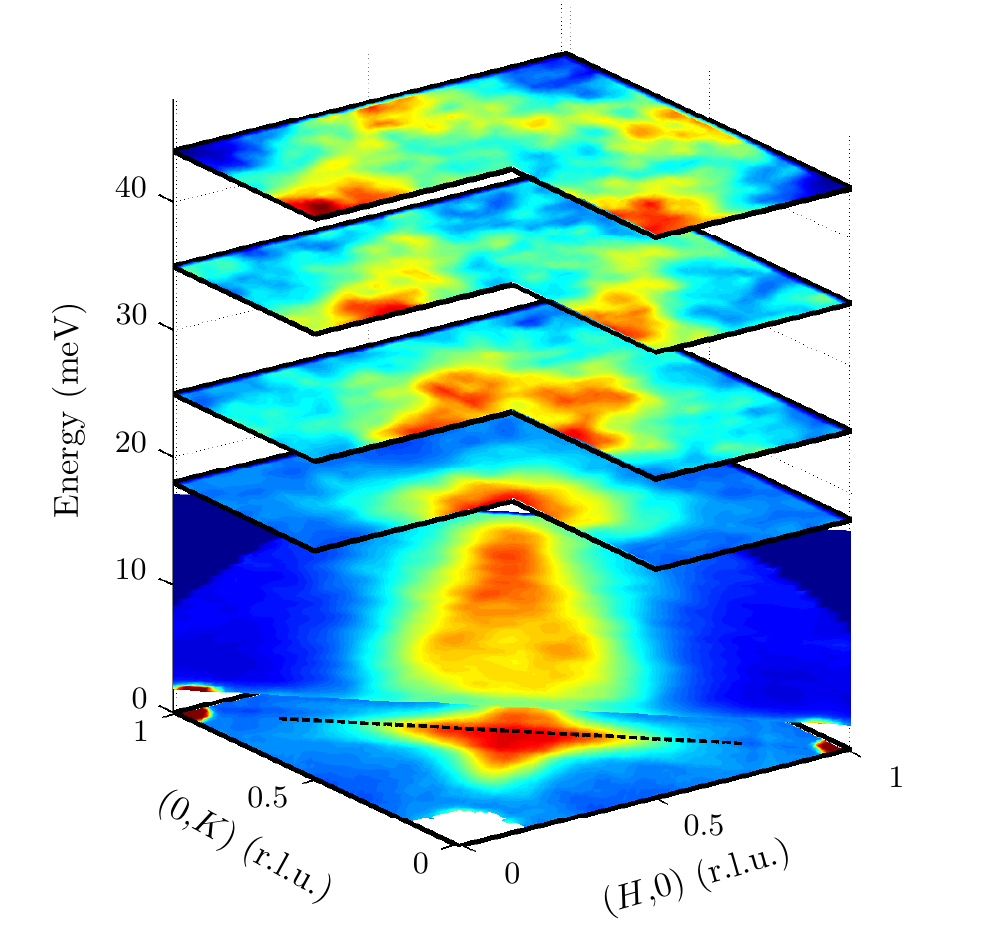}
\caption{\label{fig:Hour-Glass} (Color online)  Hour-glass magnetic spectrum of \LSCoOquart\ measured on the MERLIN spectrometer. Data below and above \SI{17}{\meV} were measured with $E_{\rm{i}} = 20$ and 60\,meV, respectively. The intensities in each plane have been scaled by different factors to help visualize the overall spectrum.}
\end{figure}

Figure \ref{fig:Hour-Glass}(a) provides an overview of our neutron scattering data on the magnetic spectrum of \LSCoOquart. The key features of the data in order of increasing energy are, \begin{inparaenum}[(i)] \item low energy branches dispersing inwards from the four incommensurate wave vectors $\mathbf{Q}_{\rm{m}}$ towards ${\bf Q}_{\rm AFM}$, \item near-vertical dispersion at ${\bf Q}_{\rm AFM}$ between about 10 and 18\,meV, and \item an outward dispersion of intensity away from ${\bf Q}_{\rm AFM}$ above about 18\,meV, with a four-peak pattern that is rotated by \ang{45} with respect to the four low energy peaks\end{inparaenum}. The spectrum has all the characteristic feature of the hour-glass spectrum, and is very similar to that observed in La$_{1.67}$Sr$_{0.33}$CoO$_4$. The main difference is that the intensity is broader in $\bf Q$, consistent with the shorter magnetic correlations in \LSCoOquart.

\begin{figure}
\includegraphics[width=\columnwidth]{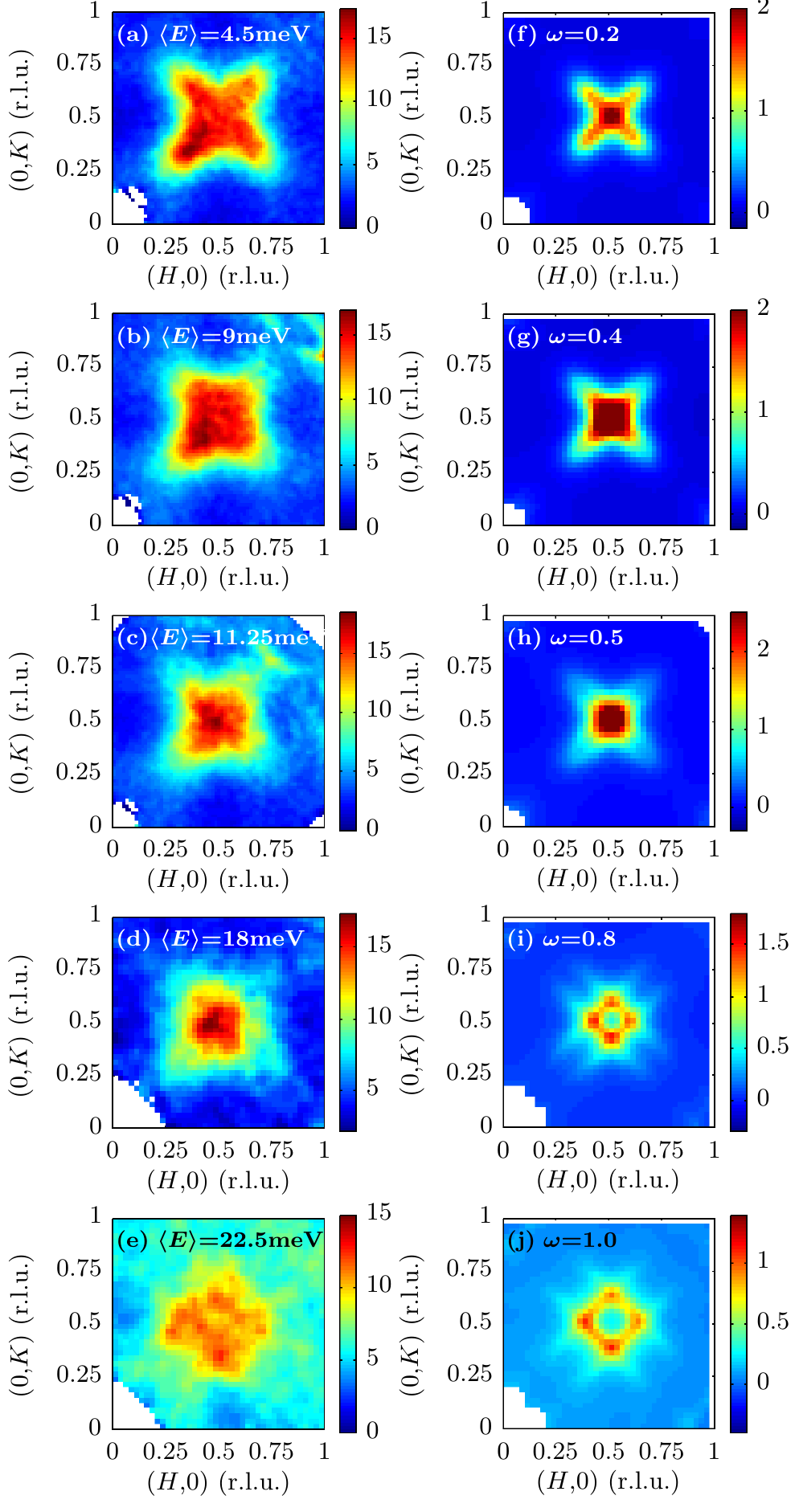}
\caption{\label{fig:Hour-Glass_constE} (Color online) (a)--(e) Constant-energy slices through the magnetic spectrum of \LSCoOquart. Data was measured on the MERLIN spectrometer using $E_{\rm{i}}=20$\,meV (a-c) and $E_{\rm{i}}=60$\,meV (d,e).  The average energy $\langle E \rangle$ of each slice is indicated. (f)--(j) DCSG model simulations at the energies indicated expressed in terms of $JS$. The simulations have been performed with $J^{\prime}/J=0.05$, and $JS = 22.5$\,meV was chosen to obtain an approximate match between the simulations and data.  The simulated charge-disordered ground state was characterized by a correlation length $\xi_{\rm{C}}^{\parallel}=5a$. The squared magnetic form factor $f^2(Q)$ and the orientation factor for in-plane magnetic fluctuations (see: Appendix~\ref{app:polarisation_factor}) have been included in the simulations. }
\end{figure}

\begin{figure}
\includegraphics[width=\columnwidth]{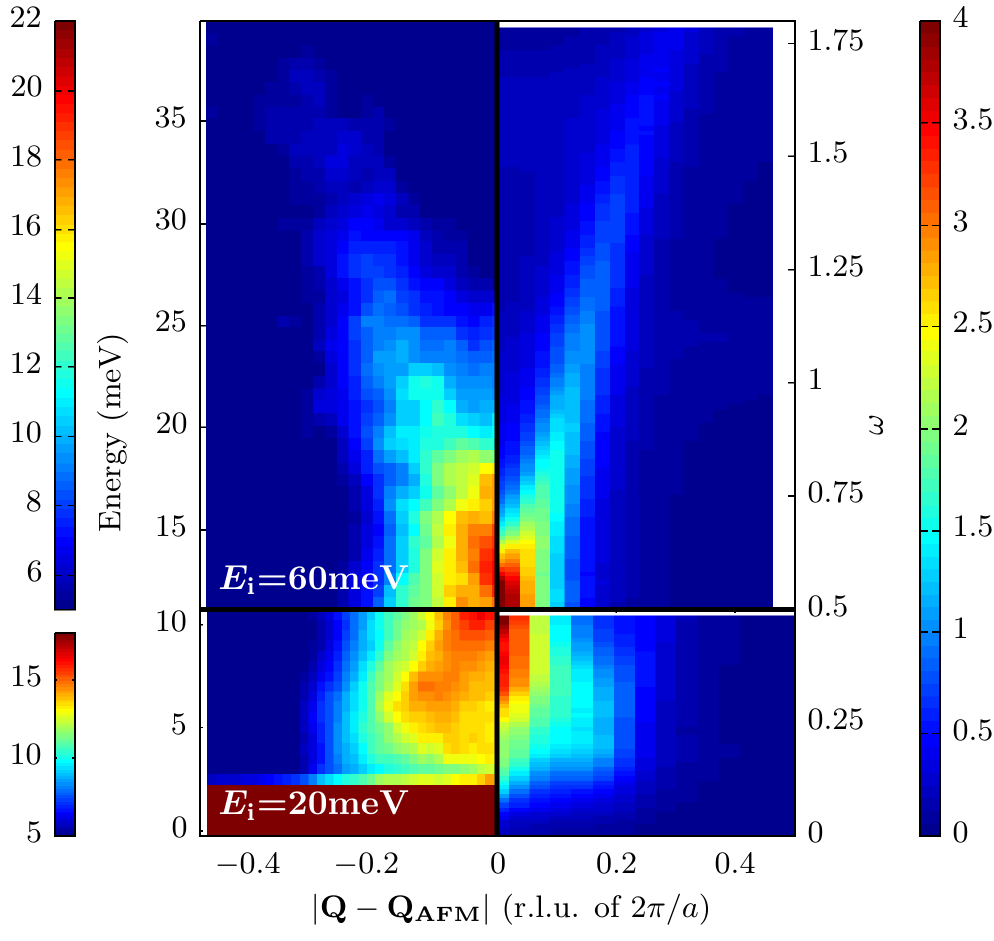}
\caption{\label{fig:Hour-Glass_disp} (Color online) Dispersion of the scattering intensity in \LSCoOquart. Lower panels ($E<11.25$\,meV, $\omega < 0.5 JS$) shows the dispersion along the ${\bf Q}=(H,H)$ direction. Upper panel ($E>11.25$\,meV, $\omega > 0.5 JS$) shows the dispersion along the $(H,0)$ direction. The left half shows neutron scattering data collected on the MERLIN spectrometer. Two incident energies were used, $E_{\rm{i}} = 20$\,meV for data in the lower panel, and  $E_{\rm{i}} = 60$\,meV in the upper panel. DCSG model simulations are plotted on the right half of the figure. The simulations have been performed with $J^{\prime}/J=0.05$, and $JS = 22.5$\,meV. The simulated charge-disordered ground state was characterized by a correlation length $\xi_{\rm{C}}^{\parallel}=5a$.  The squared magnetic form factor $f^2(Q)$ and the orientation factor for in-plane magnetic fluctuations (see: Appendix~\ref{app:polarisation_factor}) have been included in the simulations.}
\end{figure}

Figure~\ref{fig:Hour-Glass_constE}(a--e) show a series of constant-energy slices at increasing energies through the spectrum to emphasize the distribution in intensity in the $(H,K)$ plane, and the left half of Fig.~\ref{fig:Hour-Glass_disp} is an energy--wave vector slice which shows dispersion of the intensity and, in particular, the energy broadening at ${\bf Q}_{\rm AFM}$ (i.e.~the saddle point of the hour-glass, centered at $E_{\rm s}\approx 12$\,meV). In the lower part of this figure ($E < 11.25$\,meV) the wave vector on the horizontal axis is along the $(H,H)$ direction through, and in the upper part it is along the $(H,0)$ direction. In both parts, the center of the horizontal axis is at ${\bf Q}_{\rm AFM}$ and plots the magnitude of $\bf Q$ in units of $2\pi/a$.

\section{\label{ref:Analysis and Discussion} Analysis and Discussion}

\begin{figure}
\includegraphics[width=\columnwidth]{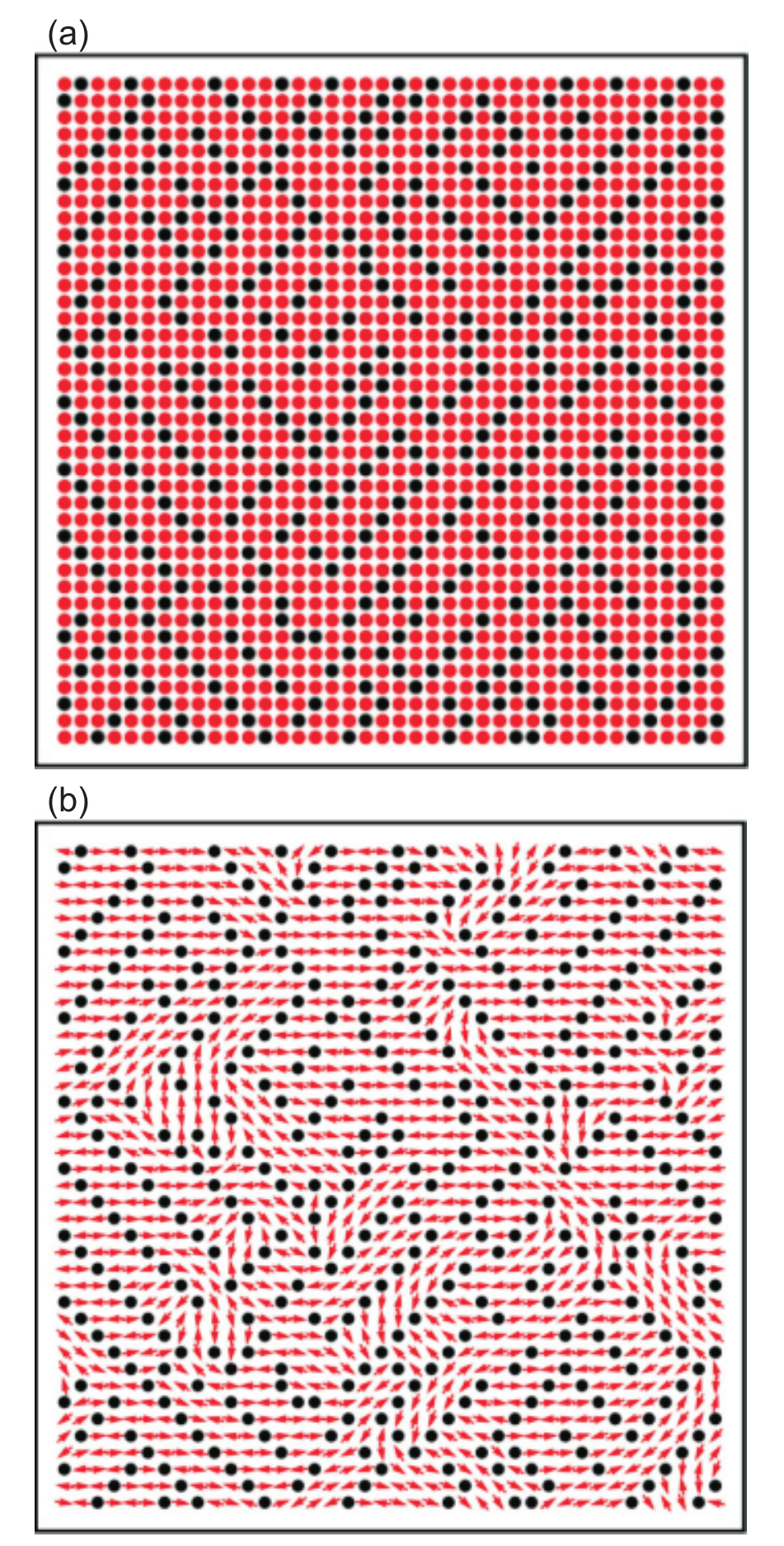}
\caption{\label{fig:charge_spin_real} (Color online) (a) Disordered charge configuration with $\xicpa\approx 5a$ and $\xicpe\approx 3.5a$, obtained from the Ising model for $N=40$. Black (red) circles correspond to Co$^{3+}$ (Co$^{2+}$) ions. (b) Corresponding real-space spin configurations in classical ground states of $\HSP$ with $\xi_{\rm{M}}^{\parallel}\approx 4.5a$ and $\xi_{\rm{M}}^{\perp}\approx 2.5a$. The dots show the non-magnetic sites, while the arrows show the $x$ and $y$ components of the ${\bf S}_{i}$ (the $z$ components are tiny due to the strong anisotropy caused by $\delta$).}
\end{figure}

The observations of signature spin-glass behaviour in the magnetization results and a very broad INS spectrum indicate that disorder is central to any theoretical description of magnetic excitations in \LSCoOquart.  In previous studies, the hour-glass spectrum was shown to arise from the combined effects of quasi-one-dimensional AFM correlations and disorder.\cite{BoothroydNature2011,UlbrichPRL2012}  The measured hour-glass spectra have been compared with spectra calculated by spin wave theory (SWT) for ideal stripes, and disorder has been included by numerical broadening. Recently, however, Andrade \textit{et al.}\cite{AndradePRL2012} developed a microscopic model which describes a system with short-range stripe order as a disordered cluster spin glass (DCSG). The model was applied to the case of \LSCoOthird, and the calculated magnetic spectrum was found to give a very good description of the observed hour-glass spectrum in this material, and captured salient features of the data that could not be satisfactorily accounted for by numerically-broadened SWT. We now extend this DCSG model to the case of \LSCoOquart.

\subsection{\label{subsec:DCSG Model} DCSG Model}

To describe disordered stripes with short-range order, we use the model proposed in Ref.~\onlinecite{AndradePRL2012} (which itself was adapted from a model proposed for fluctuating stripes in cuprates in Ref.~\onlinecite{VojtaPRL2006}). We start by modelling stripes of Co$^{2+}$ and Co$^{3+}$ ions. We assume the charge order to be static on the time scale of the magnetic fluctuations, such that the spin sector can be analyzed for a set of frozen charge configurations.\footnote{Recent experimental results, Ref.~\onlinecite{LancasterArXiv2013}, actually shows that the charge quenching remains incomplete even below the charge ordering temperature.} Those configurations correspond to imperfect stripes. We construct an Ising model for variables $n_{i}$ on the sites $i$ of a square lattice, where $n_{i}=0$ refers to a Co$^{2+}$ and $n_{i}=1$ to a Co$^{3+}$ ion. The model is chosen such that it has perfect stripe configurations with charge ordering wave vector ${\bf Q}_{\rm C}=(0.5\pm 0.25,0.5\mp 0.25)$ and ${\bf Q}_{\rm C}=(0.5\pm 0.25,0.5\pm 0.25)$ as ground states at fixed filling $\langle n_{\rm h}\rangle=1/4$. Monte-Carlo (MC) simulations of this model on $Na\times Na$ lattices are used to generate charge configurations away from this ordered state, i.e., with well-defined short-range order.\footnote{Following Ref.~\onlinecite{AndradePRL2012} we consider only short range interactions. An approach including long-range interactions was recently discussed in Ref.~\onlinecite{RademakerArXiv2013}.} Representative results are in Fig.~\ref{fig:charge_spin_real}(a) showing multiple stripe domains and various types of defects. We characterize these configurations by their correlation lengths $\xicpa$ ($\xicpe$) parallel (perpendicular) to the stripe direction, in the same fashion as defined for the magnetic correlation length as in Fig.~\ref{fig:elastic_data}(b) and (c).

To model the magnetism, we place localized $S=1/2$ spins on the Co$^{2+}$ sites of a 2D disordered stripe configuration as in Fig.~\ref{fig:charge_spin_real}(a). We assume the Co$^{3+}$ sites are spinless, as found in previous studies of La$_{2-x}$Sr$_{x}$CoO$_4$ (Refs.~\onlinecite{HollmannNJPhys2008,HelmePRB2009,BoothroydNature2011}). The spins are assumed to interact via Heisenberg exchanges as in Fig.~\ref{fig:elastic_data}(c)
\begin{equation}
\HSP=\sum_{\langle i,j \rangle}\sum_{\alpha}J_{ij}^{\alpha}S_{i}^{\alpha}S_{j}^{\alpha}.\label{eq:h_spin}
\end{equation}
The first sum runs over the lattice sites with $n_{i}=0$ according to a given charge configuration $\{n_{i}\}$, and $\alpha=x,y,z$. As in Ref.~\onlinecite{AndradePRL2012} we assume that the nearest-neighbor coupling $J$ has the same form as in the undoped parent compound: $J^{x}=J\left(1+\epsilon\right)$, $J^{y}=J$, and $J^{z}=J\left(1-\delta\right)$. The parameters $\delta$ and $\epsilon$ control the spin anisotropy with $\delta=0.28$ and $\epsilon=0.013$. Because the coupling across a Co$^{3+}$ ion, $J'$, is considerably smaller than $J$, we assume it to be isotropic. Interlayer couplings are neglected. For imperfect charge order, $\HSP$ describes frustrated magnetism due to the antiphase-domain-wall property of stripes. It is precisely this combination of disorder and frustration which leads to a DCSG.\cite{AndradePRL2012}

To access the spin dynamics, we first determine locally stable classical states of $\HSP$ for a given charge configuration $\{n_{i}\}$, taking into account the glassy nature of the problem.\cite{AndradePRL2012} An example of such ground state is illustrated in Fig.~\ref{fig:charge_spin_real}(b), which nicely shows clusters with local AFM order and essentially random relative orientations. We then calculate the excitation spectrum of $\HSP$ using linear spin-wave theory on finite lattices. Deviations from a classical state are represented by non-interacting bosons modelling Gaussian magnetic fluctuations. The spectrum is described by the total scattering function,
\begin{equation}
S^{\rm sim}({\bf Q},E)=\Big[\sum_{\alpha}S^{\alpha\alpha}({\bf Q},E)\Big]_{\rm{avg}}
\label{eq:DCSG_SQE}
\end{equation}
where $S^{\alpha\alpha}({\bf Q},E)$ is the same as that defined for Eq.~(\ref{eq:Saa}), with now $M^{\alpha}\left(\mathbf{Q}\right)$ being the Fourier-transformed spin operator. The neglect of orbital angular momentum in the model means that intensities calculated from the $S^{\alpha\alpha}$ functions are not quantitatively comparable with the real system. Where necessary, we have scaled the intensity of the simulated spectra to match the experimental data. We express energies in multiples of $JS$, i.e.~$E = \omega JS$. The $[...]_{avg}$ notation expresses that we average the $S^{\alpha\alpha}({\bf Q},E)$ summation over $80$ spin states obtained from $40$ charge configurations.

\subsection{\label{subsec: Comparison of DCSG model results to measured excitation spectrum} Comparison of DCSG Model Results to Measured Excitation Spectrum}

In order to assess the results of these simulations we have plotted 2D slices alongside the experimental data in Figs.~\ref{fig:Hour-Glass_constE} and \ref{fig:Hour-Glass_disp}. The simulated spectrum $S^{\rm sim}({\bf Q},E)$ has been multiplied by $f^2(Q)$, the squared dipole form factor of Co$^{2+}$ (Ref.~\onlinecite{NeutronDataBooklet}) and by a weighting factor that corresponds to the $\bf Q$-dependent orientation factor in Eq.~(\ref{eq:S(Q,E)}) for the case when the magnetic fluctuations are constrained to a plane -- see Appendix~\ref{app:polarisation_factor}. The justification for neglecting the out-of-plane fluctuations is that the large planar anisotropy in \LSCoOquart\ and the particular geometry employed in the experiments make both the data and $S^{\rm sim}({\bf Q},E)$ rather insensitive to out-of-plane fluctuations in the energy range covered in Figs.~\ref{fig:Hour-Glass_constE} and \ref{fig:Hour-Glass_disp}, i.e.~$S^{zz}({\bf Q},E) \ll \{S^{xx}({\bf Q},E)+S^{yy}({\bf Q},E)\}$. This approximation is discussed in more detail in Appendix B.

The simulations have been scaled such that $JS=22.5$\,meV. This value was found to give a good match between the positions of the main hour-glass features in the experimental and simulated data. The inter-stripe exchange parameter, $J^{\prime}$, was fixed to $J^{\prime}/J=0.05$. This ratio was obtained from SWT based on perfect stripe order in \LSCoOthird\ (Ref.~\onlinecite{BoothroydNature2011}), and it also gave a good description of the hour-glass dispersion calculated with the DCSG model.\cite{AndradePRL2012} We will assume $J^{\prime}/J$ remains unchanged for \LSCoOquart.  The level of disorder used in the model was chosen through comparison with various simulated spectra. Figures~\ref{fig:Hour-Glass_constE} and \ref{fig:Hour-Glass_disp} show simulations characterized by $\xi_{\rm{C}}^{\parallel}=5a$. This value was used as it reproduces the main features in the spectrum.

The DCSG simulations qualitatively reproduce all the features of the hour-glass spectrum outlined above and measured in \LSCoOquart. The distribution of intensity across the Brillioun zone, shown in Fig.~\ref{fig:Hour-Glass_constE}(f--j), illustrates this good match. However, there are discrepancies between the simulations and experimental data. The first is that the experimental data appears to be broader than simulations. Some (but not all) of this additional broadening is due to the spectrometer resolution, which is not included in the simulations.  The second shortcoming of the simulation is the  overestimation of intensity at the magnetic zone center, most noticeably between 5 and 15\,meV in Fig.~\ref{fig:Hour-Glass_disp}.  In the experimental data below 11\,meV, there are broad but nearly-resolved incommensurate excitations originating from the $\mathbf{Q}_{\rm{m}}$ positions which disperse inwards towards $\mathbf{Q}_{\rm{AFM}}$, whereas in the simulation, over the same energy range the intensity is always largest at $\mathbf{Q}_{\rm{AFM}}$.  This suggests that the simulation contains more regions of local AFM order than in the real system, although some of the discrepancy might be reduced if the parameters of the model were further refined against the data.

\begin{figure}
\includegraphics[width=\columnwidth]{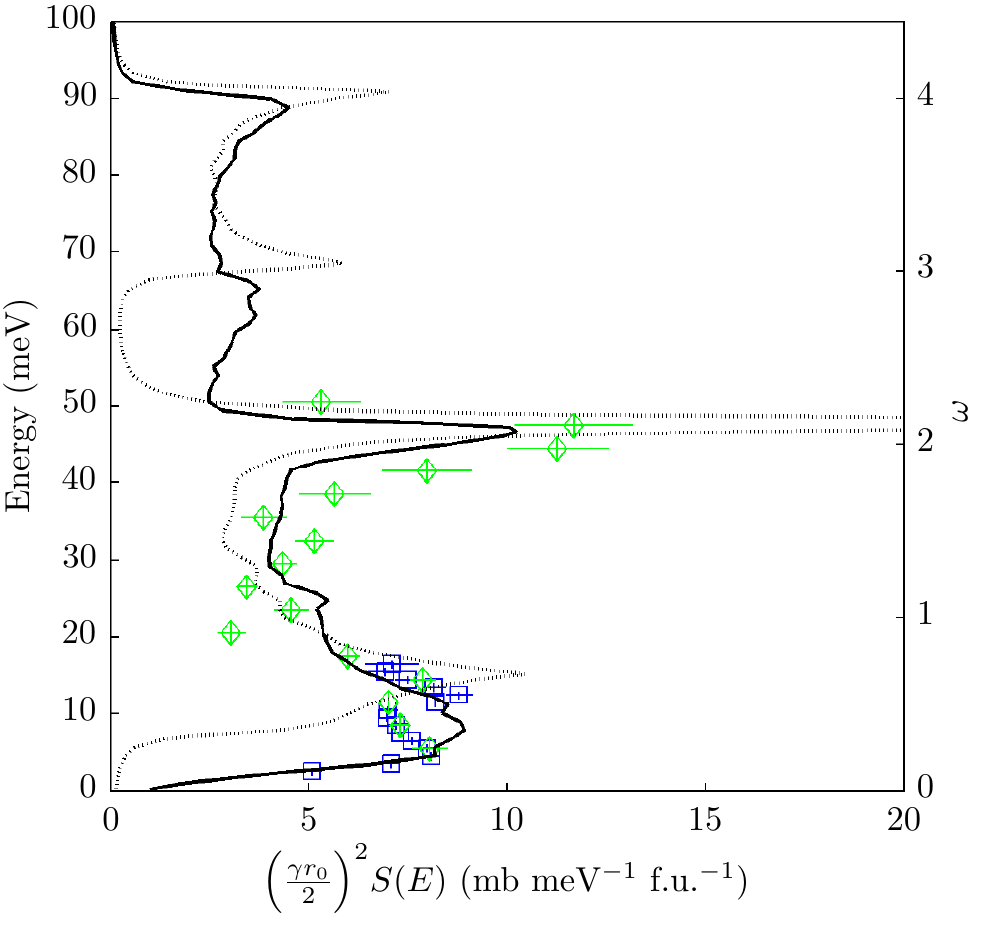}
\caption{\label{fig:chi_w} (Color online) Momentum-averaged scattering of \LSCoOquart. Experimental points are calculated by the method described in the text from the INS spectra recorded on MERLIN.  Blue (green) points indicate values fitted to data measured with $E_{\rm{i}} = 20$\,meV (\SI{60}{\meV}).  Scaled DCSG simulations of the in-plane magnetic scattering for $\xi_{\rm{C}}^{\parallel}=5a$ disordered stripes (solid line) and $\xi_{\rm{C}}^{\parallel}=\infty$ perfect stripe order (dotted line) are also plotted. Both simulations are performed with $JS=22.5$\,meV and $J'/J=0.05$.}
\end{figure}

Figure~\ref{fig:chi_w} contains momentum-averaged spectra related to the momentum-averaged partial scattering functions
\begin{equation}
S^{\alpha\alpha}(E) = \frac{\int S^{\alpha\alpha}({\bf Q},E)\,{\rm d}{\bf Q}}{\int {\rm d}{\bf Q}}.
\label{eq:S(E)}
\end{equation}
These functions are proportional to the local susceptibilities, and are a measure of the densities of magnetic modes for each magnetic polarization direction. The actual experimental quantity plotted in Fig.~\ref{fig:chi_w} is $(\gamma r_0/2)^2S^{\rm expt}(E)$, the momentum-averaged intensity corrected for the squared magnetic form factor $f^2(Q)$ and the factor $\{(1-\hat{Q}_a^2)+(1-\hat{Q}_b^2)\}/2$ corresponding to the orientation factor for in-plane fluctuations -- see Appendix A. As mentioned earlier, and detailed in Appendices A and B, the experiment is relatively insensitive to out-of-plane magnetic fluctuations, so
\begin{equation}
 S^{\rm expt}(E) \approx S^{xx}(E) + S^{yy}(E).
\end{equation}
The data points in Fig.~\ref{fig:chi_w} are obtained from the integrals of the pattern of four bi-variant Lorentzians previously used to describe the elastic diffuse scattering and now fitted to a series of constant-energy slices. The solid line shows the in-plane magnetic scattering $S^{xx}(E) + S^{yy}(E)$ calculated from the DCSG model for \LSCoOquart\ with $\xi_{\rm{C}}^{\parallel}=5a$, and the dotted line is the calculation for perfectly ordered period-4 stripes ($\xi_{\rm{C}}^{\parallel}=\infty$). Both simulations were performed with $J'/J = 0.05$, and the simulated intensities have been scaled so that the DCSG curve matches the experimental data at low energies. A comparison of the two simulated curves shows that as well as generally broadening the features of the spectrum, disorder also has the effect of shifting the saddle-point feature down in energy, from about 15\,meV in the case of ideal stripes to approximately 12\,meV for disordered stripes. This downshift of the saddle point was also observed in the DCSG simulations of \LSCoOthird\ (Ref.~\onlinecite{AndradePRL2012}).

The DCSG simulation gives a reasonable description of the features observed in the experimental data, which extends up to $\sim$50\,meV. The simulations show additional features above $\sim$50\,meV extending up to a maximum energy of $4JS$, which is the expected band width for period-4 stripes (because spins in the middle of an AFM stripe are coupled to four nearest neighbors by an exchange interaction $J$ --- see Fig.~\ref{fig:elastic_data}). These high-energy features are due to magnon bands associated with the AFM order of period-4 stripes\cite{KrugerPRB2003, CarlsonPRB2004} together with modes arising from local spin configurations present due to disorder.  Unfortunately the signal was too weak in our measurement to obtain useful data above $\sim$50\,meV. Measurements with sufficient sensitivity to probe these high-energy modes could provide useful information on local AFM correlations in \LSCoOquart.

\section{Conclusions}

We have observed evidence of short-range stripe order and the distinctive hour-glass magnetic spectrum in the layered cobaltate \LSCoOquart. The results show that the stripe phase in \LSCoOfam\ extends to lower doping than represented in the phase diagram proposed by Cwik \textit{et al.} in Ref.~\onlinecite{CwikPRL2009}. The results show that the hour-glass spectrum is robust to quite considerable amounts of disorder, characterized in \LSCoOquart\ by magnetic correlation lengths of $\xi_{\rm{M}}^{\parallel}=7.0\rm{\AA}$ and $\xi_{\rm{M}}^{\perp}=3.5\rm{\AA}$ parallel and perpendicular to the stripes.

The disordered cluster spin glass ground-state (DCSG), proposed in Ref.~\onlinecite{AndradePRL2012} to describe the disordered period-3 stripe phase found in \LSCoOthird, qualitatively reproduces all the features of the measured hour-glass spectrum of \LSCoOquart.  Investigations of the magnetic excitations at higher energies than reported here could lead to a fuller understanding of the DCSG ground-state in \LSCoOquart.

\begin{acknowledgments}
This work was supported by the UK Engineering \& Physical Sciences Research Council, and by the DFG through grant Nos. FOR 960 and GRK 1621.
\end{acknowledgments}

\appendix
\section{\label{app:polarisation_factor} Scattering from in-plane fluctuations}

The neutron inelastic scattering cross-section, Eq.~(\ref{eq:S(Q,E)}), contains the sum of partial scattering functions weighted by factors which filter out the components of the magnetic fluctuations along the scattering vector $\bf Q$:
\begin{equation}
S^{\rm INS}({\bf Q},E) = \sum_{\alpha}(1-\hat{Q}_{\alpha}^2)S^{\alpha\alpha}({\bf Q},E).
\label{eq:A1}
\end{equation}
The DCSG simulations, on the other hand, sum the partial scattering functions without the weighting factors --- Eq.~(\ref{eq:DCSG_SQE}):
\begin{equation}
S^{\rm sim}({\bf Q},E) = \sum_{\alpha}S^{\alpha\alpha}({\bf Q},E).
\label{eq:A2}
\end{equation}
In general, $S^{\rm INS}$ and $S^{\rm sim}$ are not proportional to one another and so cannot be compared quantitatively. However, as we show here, $S^{\rm INS}$ and $S^{\rm sim}$ are proportional to one another for the special case where the magnetic fluctuations are constrained to a plane and are domain-averaged. As shown in Appendix~\ref{app:Effect of Out-of-Plane Fluctuations}, this case applies to \LSCoOquart\ due to the strong planar anisotropy and the particular experimental configuration used here.

We denote the easy direction for the spins within the plane to be $x$, and we assume an equal probability for $x$ to be parallel to the $a$ and $b$ axes of the tetragonal crystal lattice. Setting $S^{zz}({\bf Q},E) = 0$ in Eqs.~(\ref{eq:A1}) and (\ref{eq:A2}) and averaging over the two domains, we find,
\begin{align}
& \left \langle S^{\rm INS}({\bf Q},E) \right \rangle \nonumber \\
&\qquad =\frac{1}{2}\left[(1-\hat{Q}_{a}^2) S^{xx}({\bf Q},E) + (1-\hat{Q}_{b}^2) S^{yy}({\bf Q},E)  \right] \nonumber \\
&\qquad +\frac{1}{2}\left[(1-\hat{Q}_{a}^2) S^{yy}({\bf Q},E) + (1-\hat{Q}_{b}^2) S^{xx}({\bf Q},E)  \right] \nonumber \\[8pt]
& \qquad =\frac{1}{2}\left[(1-\hat{Q}_{a}^2)+(1-\hat{Q}_{b}^2) \right] \left[S^{xx}({\bf Q},E) + S^{yy}({\bf Q},E) \right] \nonumber \\[8pt]
& \qquad = \frac{1}{2} (1+\hat{Q}_{c}^2) S^{\rm sim}({\bf Q},E).
\label{eq:A3}
\end{align}

Here, $\left \langle \ldots \right \rangle$ denotes the domain average in the real crystal, and $\hat{Q}_{a}$, $\hat{Q}_{b}$, $\hat{Q}_{c}$ are the components of the unit vector $\hat{\bf Q}$ along the tetragonal crystal axes.  Hence, for the case of in-plane fluctuations, $S^{\rm INS}$ and $S^{\rm sim}$ are proportional to one another, and can be compared directly after correction for the prefactor $\frac{1}{2} (1+\hat{Q}_{c}^2)$.

\section{\label{app:Effect of Out-of-Plane Fluctuations} Effect of Out-of-Plane Spin Fluctuations}

\begin{figure}[h]
\includegraphics[width=\columnwidth]{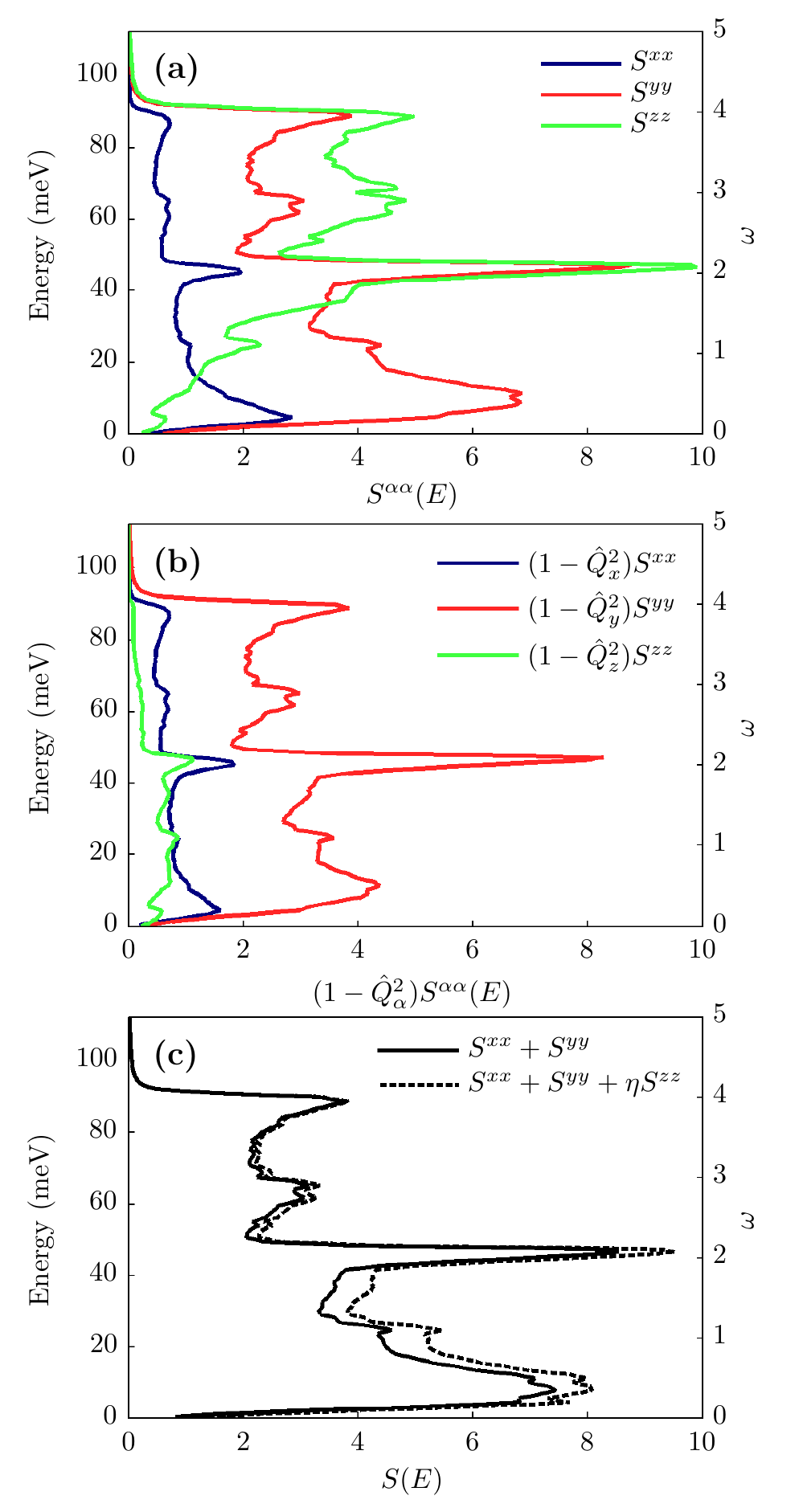}
\caption{\label{fig:Sw_xyz}(Color online) (a) Momentum-averaged partial scattering functions $S^{xx}$ (blue line), $S^{yy}$ (red line) and $S^{zz}$ (green line), simulated for the case where $\xi_{\rm{C}}^{\parallel}=5a$, $J=22.5$~meV and $J'/J=0.05$. (b) The partial scattering functions weighted by the orientation factors in the INS cross-section. The orientation factors are calculated from the variation of $\bf Q$ with energy in the time-of-flight spectrum, with the in-plane component of $\bf Q$ fixed at $(0.5,0.5)$. The neutron incident energy was $E_{\rm{i}}=60$~meV, as used in the experiment.  (c) Comparison of the scattering from the in-plane fluctuations $S^{xx}+S^{yy}$ (solid black line) with the actual combination of scattering functions present in the experimental INS spectrum, $S^{xx}(E) + S^{yy}(E) + \eta S^{zz}(E)$ with $\eta = 2(1-\hat{Q}_{c}^2)/(1+\hat{Q}_{c}^2)$ (dashed black line). The in-plane fluctuations are seen to dominate the experimental spectrum over the entire bandwidth.}
\end{figure}

Here we use results from the DCSG model simulations to demonstrate that our INS measurements are relatively insensitive to out-of-plane magnetic fluctuations.

Fig.~\ref{fig:Sw_xyz}(a) displays the individual components $S^{xx}(E)$, $S^{yy}(E)$ and $S^{zz}(E)$ of the momentum-averaged scattering function calculated from the DCSG model with $\xi_{\rm{C}}^{\parallel}=5a$. The majority of the signal below $\sim 20$\,meV is seen to be from the in-plane fluctuations, $S^{xx}(E) + S^{yy}(E)$, as expected for the strong XY-like magnetic anisotropy in \LSCoOquart. The $S^{zz}(E)$ component is largest for $E>35$~meV.

In Fig.~\ref{fig:Sw_xyz}(b) we show the functions $(1-\hat{Q}_{\alpha}^2)S^{\alpha\alpha}({\bf Q},E)$ which appear in the INS cross-section, Eq.~(\ref{eq:S(Q,E)}). Now it can be seen that the in-plane fluctuations dominate over the entire bandwidth of the measured INS spectrum. This is because in the time-of-flight method, when the incident beam is parallel to the $c$-axis the component of $\bf Q$ parallel to $c$ at small scattering angles increases with increasing energy, and so the orientation factor $(1-\hat{Q}_{c}^2)$ decreases with increasing energy. Hence, for energies above $\sim 20$\,meV, where $S^{zz}(E)$ becomes important, the orientation factor suppresses $S^{zz}(E)$ relative to $S^{xx}(E)$ and $S^{yy}(E)$.

Finally, in Fig.~\ref{fig:Sw_xyz}(c) we show the calculated spectrum of in-plane fluctuations $S^{xx}(E) + S^{yy}(E)$ together with the combination $S^{xx}(E) + S^{yy}(E) + \eta S^{zz}(E)$, where $\eta = 2(1-\hat{Q}_{c}^2)/(1+\hat{Q}_{c}^2)$ determines the proportion of $S^{zz}(E)$ which appears in the experimental spectrum ($\eta$ is the ratio of out-of-plane to in-plane orientation factors, see Eq.~(\ref{eq:A3})). A comparison of these two curves shows that the $S^{zz}(E)$ component is at most 15\% of the total measured intensity.

The results shown in Figs.~\ref{fig:Sw_xyz} confirm that the experimental spectrum is dominated by scattering from in-plane magnetic fluctuations.

\bibliographystyle{apsrev4-1}
\bibliography{LS025CoO_HourGlass_bib}

\end{document}